\documentstyle[11pt,newpasp,twoside]{article}
\def\etal{{et\thinspace al.}\ }
\begin{document}
\title{NLTE Radiative Transfer and Model Atmospheres of Hot Stars}
\author{Klaus Werner}
\affil{Institut f\"ur Astronomie und Astrophysik, Universit\"at T\"ubingen, 
Sand 1, D-72076 T\"ubingen, Germany}

\begin{abstract}
We describe our method to construct line blanketed NLTE model atmospheres for hot
stars. We employ the Accelerated Lambda Iteration and use statistical methods
to deal with metal line blanketing.
\end{abstract}

\section{Introduction}

Stellar atmospheres are open systems and thus cannot be in thermodynamic
equilibrium (TE). The ``Local Thermodynamic Equilibrium'' (LTE) is a working
hypothesis which assumes TE not for the atmosphere as a whole but for small
volume elements. As a consequence, the atomic population numbers are depending
only on the local (electron) temperature and electron density via the
Saha-Boltzmann equations. Computing models by replacing the Saha-Boltzmann
equations by the rate equations (statistical equilibrium) are called non-LTE
(or NLTE) models. NLTE calculations are more costly than LTE calculations,
however, it is hard to predict if NLTE effects are important in a specific
problem. Generally, NLTE effects are large at high temperatures and low
densities, which implies intense radiation fields hence frequent radiative
processes and less frequent particle collisions which tend to enforce LTE
conditions.

We will restrict ourselves here to the classical model atmosphere problem,
i.e.\ the solution of the radiation transfer equations assuming hydrostatic,
radiative and statistical equilibrium. The numerical problem going from LTE to
realistic NLTE models has been solved only recently and is the topic of this
paper. This was achieved by the development of new numerical techniques for
model construction and on the availability of atomic data for many species. The
replacement of the Saha-Boltzmann equations by the atomic rate equations
requires a different numerical solution technique, otherwise metal opacities
cannot be accounted for at all. Such techniques were developed with big success
during the last decade, triggered by important papers by Cannon (1973) and
Scharmer (1981). The Accelerated Lambda Iteration (ALI) is at the heart of of
this development. Combined with statistical methods we are finally able to
compute metal line blanketed NLTE models with a very high level of
sophistication.

\section{Overview of problem and solution method\label{eins2}}

We assume plane-parallel geometry, which is well justified for most stars
because the atmospheres are thin compared to the stellar radius. The only
parameters which characterize uniquely such an atmosphere are the effective
temperature (\hbox{$T_{\rm eff}$}), which is a measure for the amount of energy
transported through the atmosphere per unit area and time, the surface gravity
($g$), and the chemical composition. To construct model atmospheres we have to
solve simultaneously a set of equations that is highly coupled and non-linear.
Because of the coupling, no equation is determining uniquely a single quantity
-- all equations determine a number of state parameters. However, each of them
is usually thought of as determining a particular quantity. These equations are:

\begin{itemize}
\item The radiation transfer equations which are solved for the (angular) mean
intensities $J_i, i=1,\ldots,NF$, on a pre-chosen frequency grid comprising
$NF$ points. The formal solution is given by $J=\Lambda S$, where $S$ is the
source function (Eq.\,\ref{source}). Although $\Lambda$ is
written as an operator, one may think of $\Lambda$ as a {\em process} of
obtaining the mean intensity from the source function.
\item The hydrostatic equation which determines the total particle
density $N$.
\item The radiative equilibrium equation from which the temperature $T$ follows.
\item The particle conservation equation, determining the electron density 
$n_e$.
\item The statistical equilibrium equations which are solved for the population
densities $n_i, i=1,\ldots,NL$  of the atomic NLTE levels.
\item The definition equation for a fictitious massive particle density $n_H$
which is introduced for a convenient representation of the solution procedure.
\end{itemize}
This set of equations has to be solved at each point $d$ of a grid comprising
$ND$ depth points. Thus we are looking for solution vectors
\begin{equation}\label{psi1}
\mbox{\boldmath$\psi$}_d' = (n_1,\ldots,n_{NL}, n_e, T, n_H, N,
J_1,\ldots,J_{NF}) , \quad d=1, \ldots ,ND .
\end{equation}
The Complete Linearization (CL) method (Auer \& Mihalas 1969) solves this set by
linearizing the equations with respect to all variables. The basic advantage of
the ALI (or ``operator splitting'') method is that it allows to eliminate at
the outset the explicit occurrence of the mean intensities $J_i$ from the
solution scheme by expressing these variables by the current, yet to be
determined, occupation densities and temperature. This is accomplished by an
iteration procedure which may be written as (suppressing indices indicating
depth and frequency dependency of variables):
\begin{equation}\label{ali}
J^n=\Lambda^{\star}S^n+(\Lambda-\Lambda^{\star})S^{n-1} .
\end{equation}
This means that the actual mean intensity at any iteration step $n$ is computed
by applying an approximate lambda operator (ALO) $\Lambda^{\star}$ on the
actual (thermal) source function $S^n$ plus a correction term that is computed
from quantities known from the previous iteration step. This correction term
includes the exact lambda operator $\Lambda$ which guarantees the exact
solution of the radiation transfer problem in the limit of convergence:
$J=\Lambda S$. The use of $\Lambda$ in Eq.\,\ref{ali} only indicates that a
formal solution of the transfer equation is performed but in our application
the operator is not constructed explicitly. Instead we employ the Feautrier
solution scheme (Mihalas 1978) or a short characteristic method (Olson \&
Kunasz 1987) to solve the transfer equation that is set up as a differential
equation. The resulting set of equations for the reduced solution vectors
\begin{equation}\label{psi2}
\mbox{\boldmath$\psi$}_d = (n_1,\ldots,n_{NL}, n_e, T, n_H, N) , \quad
d=1,\ldots,ND
\end{equation}
is of course still non-linear. The solution is obtained by linearization and
iteration which is performed either with a usual Newton-Raphson iteration or
other methods (Sect.\,\ref{kanto}). First model atmosphere calculations with
the ALI method were performed by Werner (1986).

Another advantage of the ALI method is that explicit depth coupling of the
solution vectors Eq.\,\ref{psi1} through the transfer equation can be avoided
if one restricts to diagonal (i.e.\ local) ALOs. Then the solution vectors
Eq.\,\ref{psi2} are independent from each other and the solution procedure
within one iteration step of Eq.\,\ref{ali} is much more straightforward. Depth
coupling is provided by the correction term that involves the exact solution of
the transfer equation. The hydrostatic equation which also gives an explicit
depth coupling, may be taken out of the set of equations and can -- as
experience shows -- be solved in between two iteration steps of Eq.\,\ref{ali}.
Then full advantage of a local ALO can be taken. The linearized system may be
written as:
\begin{equation}
\mbox{\boldmath$\psi$}_d
=\mbox{\boldmath$\psi$}_d^0+\delta\mbox{\boldmath$\psi$}_d \ ,
\end{equation}
where $\mbox{\boldmath$\psi$}_d^0$ is the current estimate for the solution
vector at depth $d$ and $\delta\mbox{\boldmath$\psi$}_d$ is the correction
vector to be computed. Using a tri-diagonal $\Lambda^{\star}$ operator the
resulting system for $\delta\mbox{\boldmath$\psi$}_d$ is -- like in the
classical CL scheme -- of block tri-diagonal form coupling each depth point $d$
to its nearest neighbors $d\pm1$:
\begin{equation}\label{tri}
\mbox{\boldmath$\gamma$}_d\delta\mbox{\boldmath$\psi$}_{d-1}+
\mbox{\boldmath$\beta$}_d\delta\mbox{\boldmath$\psi$}_d+
\mbox{\boldmath$\alpha$}_d\delta\mbox{\boldmath$\psi$}_{d+1}={\bf c}_d .
\end{equation}
The quantities {$\mbox{\boldmath$\alpha, \beta, \gamma$}$} are ($NN\times NN$)
matrices where $NN$ is the total number of physical variables, i.e., $NN=NL+4$,
and ${\bf c}_d$ is the residual error in the equations. The solution is
obtained by the standard Feautrier scheme. As already mentioned, the system
Eq.\,\ref{tri} breaks into $ND$ independent equations
$\delta\mbox{\boldmath$\psi$}_d=\mbox{\boldmath$\beta$}_d^{-1}{\bf c}_d$
($d=1,\ldots,ND$) when a local $\Lambda^{\star}$ operator is used. The
additional numerical effort to set up the subdiagonal matrices and matrix
multiplications in the tri-diagonal case is outweighed by the faster global
convergence of the ALI cycle, accomplished by the explicit depth coupling in the
linearization procedure (Werner 1989).

The principal advantage of the ALI over the CL method becomes clear at this
point. Each matrix inversion necessary to solve Eq.\,\ref{tri} requires
$(NL+4)^3$ operations whereas in the CL method $(NL+NF+4)^3$ operations are
needed. Since the number of frequency points $NF$ is much larger than the
number of levels $NL$, the matrix inversion in the CL approach is dominated by
$NF$.

Recent developments concern the problem that the total number of atomic levels
tractable in NLTE with the ALI method described so far is restricted to the
order of 250, from experience with our model atmosphere code {\tt PRO2}. This
limit is a consequence of the non-linearity of the equations, and in order to
overcome it,  measures must be taken in order to achieve a linear system whose
numerical solution is much more stable. Such a pre-conditioning procedure has
been first applied in the ALI context by Werner \& Husfeld (1985) using the
core saturation method (Rybicki 1972). More advanced work achieves linearity by
replacing the $\Lambda$ operator with the $\Psi$ operator (and by judiciously
considering some populations as ``old'' and some as ``new'' ones within an ALI
step) which is formally defined by writing:
\begin{equation}
J_\nu=\Psi_\nu\eta_\nu\ , \qquad {\rm i.e.} \qquad
\Psi_\nu\equiv\Lambda_\nu/\chi_\nu\ ,
\end{equation}
where the total opacity $\chi_\nu$ (as defined in Sect.\,\ref{opa}) is
calculated from the previous ALI cycle. The advantage is that the emissivity
$\eta_\nu$ (Sect.\,\ref{opa}) is linear in the populations, whereas the source
function $S_\nu$ is not. Hence the new operator $\Psi$ gives the solution of the
transfer problem by acting on a linear function. This idea is based on Rybicki
\& Hummer (1991) who applied it to the line formation problem, i.e.\
restricting the set of equations to the transfer and rate equations and
regarding the atmospheric structure as fixed. Hauschildt (1993) and Hauschildt
\& Baron (1999) generalized it to solve the full model atmosphere problem. In
addition, splitting the set of statistical equations and solving it separately
for each chemical element means that now many hundreds of levels per species
are tractable in NLTE.

\section{Basic equations}

We give a short overview of the basic equations to be solved. More detailed
presentations may be found in Werner \& Dreizler (1999) or Werner \etal (2003).

\subsection{Radiation transfer}

Any numerical method requires a formal solution (i.e.\ atmospheric structure
already given) of the radiation transfer problem. The radiation transfer at any
particular depth point can be described by the following equation, formally
written for positive and negative $\mu$ (which is the cosine of the angle
between direction of propagation and outward directed normal to the surface)
separately, i.e.\ for inward and for outward directional intensities $I$ with
frequency $\nu$:
\begin{equation}\label{te}
\pm\mu \frac{\partial I_{\nu}(\pm\mu)}{\partial\tau_{\nu}}=
S_{\nu}-I_{\nu}(\pm\mu), \quad\mu\in[0,1].
\end{equation}
$\tau_\nu$ is the optical depth (which can be defined via the column mass $m$
that is used in the other structural equations by $d\tau_\nu=dm \chi_\nu/\rho$,
with the mass density $\rho$) and $S_{\nu}$ is the local source function.
Introducing the Feautrier intensity
\begin{equation}\label{udef}
u_{\nu\mu}\equiv \left(I_{\nu}(\mu)+I_{\nu}(-\mu)\right)/2
\end{equation}
we obtain the second-order form (Mihalas 1978):
\begin{equation}
\mu^2 \frac{\partial^2u_{\nu\mu}}{\partial\tau_{\nu}^2}=u_{\nu\mu}-S_{\nu}, 
\quad\mu\in[0,1].
\end{equation}
We may separate the Thomson emissivity term (scattering from free electrons,
assumed coherent, with cross-section $\sigma_e$) from the source function so
that
\begin{equation}\label{source}
S_{\nu}=S_{\nu}'+n_e\sigma_eJ_{\nu}/\chi_{\nu},
\end{equation}
where $S_{\nu}'$ is the ratio of thermal emissivity to total opacity
(Sect.\,\ref{opa}): $S_{\nu}'=\eta_{\nu}/\chi_{\nu}$. Since the mean intensity
is the angular integral over the Feautrier intensity the transfer equation
becomes
\begin{equation}\label{unm}
\mu^2 \frac{\partial^2u_{\nu\mu}}{\partial\tau_{\nu}^2}
=u_{\nu\mu}-S_{\nu}'-\frac{n_e\sigma_e}{\chi_{\nu}}\int_{0}^{1}u_{\nu\mu}\,d\mu .
\end{equation}
Thomson scattering complicates the situation by explicit angle coupling.
Assuming complete frequency redistribution in spectral lines, no explicit
frequency coupling occurs so that the parallel solution for all frequencies
enables a very efficient vectorization on the computer.

\subsection{Statistical equilibrium}

\subsubsection{Rate equations}

For each atomic level $i$ the rate equation describes the equilibrium of rates
into and rates out of this level:
\begin{equation}\label{rates}
n_i\sum_{i\neq j}^{}P_{ij}-\sum_{j\neq i}^{}n_jP_{ji}=0 .
\end{equation}
The rate coefficients $P_{ij}$ have radiative and collisional components:
$P_{ij}=R_{ij}+C_{ij}$. The radiative downward rate for example is
given by:
\begin{equation}
R_{ji}=\left(\frac{n_i}{n_j}\right)^{\star}4\pi\int_{0}^{\infty} 
\frac{\sigma_{ij}(\nu)}
{h\nu}\left(\frac{2h\nu^3}{c^2}+J_{\nu}\right)e^{-h\nu/kT}\,d\nu .
\end{equation}
Photon cross-sections are denoted by $\sigma_{ij}(\nu)$.
$({n_i}/{n_j})^{\star}$ is the Boltzmann LTE population ratio. The computation
of collisional rates is generally dependent on the specific ion or even
transition.

\subsubsection{Abundance definition equation}

The rate equation for the highest level of a given chemical species is
redundant. It is replaced by the abundance definition equation. Summation over
all levels usually includes not only NLTE levels but also levels which are
treated in LTE, according to the specification in the model atom. Denoting the
number of ionization stages of species $k$ with $NION(k)$, the number of NLTE
and LTE levels per ion with $NL(l)$ and $LTE(l)$, respectively, and the number
fraction of species $k$ with $y_k$, we can write:
\begin{equation}
\sum_{l=1}^{NION(k)}\left[\sum_{i=1}^{NL(l)}n_{kli}+
\sum_{i=1}^{LTE(l)}n_{kli}^{\star}\right]
=y_k (N-n_e) . 
\end{equation}

\subsubsection{Charge conservation}

We close the system of statistical equations by invoking charge conservation.
We denote the total number of chemical species with $\textrm{\it NATOM}$, the
charge of ion $l$ with $q(l)$ (in units of the electron charge) and write:
\begin{equation}
\sum_{k=1}^{\textrm{\scriptsize\it NATOM}\phantom{)}}\sum_{l=1}^{NION(k)}q(l)
\left[\sum_{i=1}^{NL(l)}n_{kli}+\sum_{i=1}^{LTE(l)}n_{kli}^{\star}\right]=n_e .
\end{equation}

\subsubsection{Complete statistical equilibrium equations}

We introduce a vector comprising the occupation numbers of all NLTE levels,
$\bf n$ $=(n_1, \ldots ,n_{NL})$. Then the statistical equilibrium equation is
written as:
\begin{equation}\label{amat}
\bf An=b .
\end{equation}
The gross structure of the rate matrix $\bf A$ is of block matrix form, because
transitions between levels occur within one ionization stage or to the ground
state of the next ion. The structure is complicated by ionizations into excited
levels and by the abundance definition and charge conservation equations which
give additional non-zero elements in the corresponding lines of ${\bf A}$.

\subsection{Radiative equilibrium}

Radiative equilibrium can be enforced by adjusting the temperature
stratification either during the linearization procedure or in between ALI
iterations. In the former case a linear combination of two different
formulations can be used and in the latter case the classical Uns\"old-Lucy
temperature correction procedure (Lucy 1964) is utilized. The latter is
particularly interesting, because it allows to exploit the blocked form of the
rate coefficient matrix. This will enable an economic block-by-block solution
followed by a subsequent Uns\"old-Lucy temperature correction step. On the
other side, however, this correction procedure may decelerate the global
convergence behavior of the ALI iteration.

The two forms of expressing the radiative equilibrium condition follow from the
postulation that the energy emitted by a volume element per unit time is equal
to the absorbed energy per unit time (integral form):
\begin{equation}\label{int}
\int_{0}^{\infty}\chi_{\nu}(S_{\nu}-J_{\nu})\,d\nu=0 .
\end{equation}
This formulation is
equivalent to invoking flux constancy throughout the atmosphere (differential
form) involving the nominal flux ${\cal H}$:
\begin{equation}\label{diff}
\int_{0}^{\infty}\frac{\partial}{\partial\tau_{\nu}}(f_{\nu}J_{\nu})\,d\nu-{\cal H}=0 ,
\end{equation}
where $f_{\nu}$ is the variable Eddington factor, defined as
\begin{equation}\label{eddfac}
f_{\nu}=\int_{0}^{1}\mu^2u_{\nu\mu}\,d\mu \Big/ \int_{0}^{1}u_{\nu\mu}\,d\mu
\end{equation}
and computed from the Feautrier intensity $u_{\nu\mu}$ (Eq.\,\ref{udef}) after
the formal solution. The differential form is more accurate at large depths,
while the integral form behaves numerically better at small depths. Instead of
arbitrarily selecting that depth in the atmosphere where we switch from one
formulation to the other, we use a linear combination of both constraint
equations which guarantees a smooth transition with depth, based on physical
grounds:
\begin{equation}\label{combi}
\frac{1}{\bar\kappa_J}\int_{0}^{\infty}\chi_{\nu}(S_{\nu}-J_{\nu})\,d\nu+
{\bar\Lambda_J^\star}\int_{0}^{\infty}\frac{\partial}{\partial\tau_{\nu}}(f_{\nu}J_{\nu})\,d\nu-
{\bar\Lambda_J^\star}{\cal H}-F_0=0 .
\end{equation}
For details on this equation and on our implementation of the Uns\"old-Lucy
procedure see Dreizler (2003).

\subsection{Hydrostatic equilibrium}\label{defm}

The total pressure $P$ comprises gas, radiation and turbulent pressures, so
that we write the equation for hydrostatic equilibrium as:
\begin{equation}\label{hydros}
\frac{d}{dm}P=
\frac{d}{dm}\left( NkT+\frac{4\pi}{c}\int_{0}^{\infty}f_{\nu}J_{\nu}\,d\nu
+\frac{1}{2}\rho v^2_{\rm turb}\right)=g
\end{equation}
with Boltzmann's constant $k$ and the turbulent velocity $v_{\rm turb}$. The
hydrostatic equation may either be solved simultaneously with all other
equations or separately in between iterations. The overall convergence behavior
is usually the same in both cases. If taken into the linearization scheme and a
local $\Lambda^{\star}$ operator is used then, like in the case of the
radiative equilibrium equation, explicit depth coupling enters via the depth
derivative $d/dm$. Again, solution of the linearized equations has to proceed
inwards starting at the outer boundary.

\subsection{Particle conservation and fictitious massive particle density}

The total particle density $N$ is the sum of electron density plus the
population density of all atomic states, LTE and NLTE levels: 
\begin{equation}
N=n_e+
\sum_{k=1}^{\textrm{\scriptsize\it NATOM}\phantom{)}}\sum_{l=1}^{NION(k)}
\left[\sum_{i=1}^{NL(l)}n_{kli}+\sum_{i=1}^{LTE(l)}n_{kli}^{\star}\right] .
\end{equation}
A fictitious massive particle density $n_H$ is defined by:
\begin{equation}
n_H=\sum_{k=1}^{\textrm{\scriptsize\it NATOM}\phantom{)}}m_k\sum_{l=1}^{NION(k)}
\left[\sum_{i=1}^{NL(l)}n_{kli}+\sum_{i=1}^{LTE(l)}n_{kli}^{\star}\right] .
\end{equation}
The mass of a chemical species in AMU is denoted by $m_k$. Introducing the mass
of a hydrogen atom $m_H$, we then may simply write for the material density
\begin{equation}
\rho=n_Hm_H .
\end{equation}

\subsection{Opacity and emissivity \label{opa}}

Thermal opacity and emissivity are made up by radiative bound-bound, bound-free
and free-free transitions. For each species we compute and sum up:
\begin{eqnarray}\label{chi}
\kappa_{\nu}&=&\sum_{l=1}^{NION\phantom{)}}\left[
\sum_{i=1}^{NL(l)}\sum_{j>i}^{NL(l)}\sigma_{li\rightarrow lj}(\nu)
\left(n_{li}-n_{lj}\frac{g_{li}}{g_{lj}}
e^{-h(\nu-\nu_{ij})/kT}
\right) \right. \\ 
& & + \sum_{i=1}^{NL(l)}\sum_{j>i}^{NL(l+1)}\sigma_{li\rightarrow l+1,k}(\nu)
\left(n_{li}-n_{li}^{\star}e^{-h\nu/kT}\right) \nonumber \\ 
& & + \left. n_e\sigma_{kk}(l,\nu)\left(1-e^{-h\nu/kT}\right)
\left(\sum_{i=1}^{NL(l+1)}n_{l+1,i}+\sum_{i=1}^{LTE(l+1)}n_{l+1,i}^{\star}
\right)
\right] \nonumber
\end{eqnarray}
where the total opacity includes Thomson scattering, i.e.\
$\chi_\nu=\kappa_\nu+n_e\sigma_e$, and
\begin{eqnarray}\label{eta}
\frac{\eta_{\nu}}{2h\nu^3/c^2}&=&\sum_{l=1}^{NION\phantom{)}}\left[
\sum_{i=1}^{NL(l)}\sum_{j>i}^{NL(l)}\sigma_{li\rightarrow lj}(\nu)
n_{lj}\frac{g_{li}}{g_{lj}}
e^{-h(\nu-\nu_{ij})/kT} 
\right. \\
& & +\sum_{i=1}^{NL(l)}\sum_{j>i}^{NL(l+1)}\sigma_{li\rightarrow l+1,k}(\nu)
n_{li}^{\star}e^{-h\nu/kT} \nonumber \\
& & + \left. 
n_e\sigma_{kk}(l,\nu)e^{-h\nu/kT}
\left(\sum_{i=1}^{NL(l+1)}n_{l+1,i}+\sum_{i=1}^{LTE(l+1)}n_{l+1,i}^{\star}
\right)
\right] . \nonumber
\end{eqnarray}
$\sigma_{li\rightarrow l+1,k}(\nu)$ denotes the cross-section for
photoionization from level $i$ of ion $l$ into level $k$ of ion $l+1$. The
double summation over the bound-free continua takes into account the
possibility that a particular level may be ionized into more than one level of
the next high ion. The source function used for the approximate radiation
transfer is $\eta_{\nu}/\kappa_{\nu}$, thus, excludes Thomson scattering. For
the exact formal solution of course, the total opacity $\chi_\nu$ in the
expression Eq.\,\ref{source} includes the Thomson term ($n_e\sigma_e$).

\section{The Accelerated Lambda Iteration (ALI)}

In all constraint equations described above the mean intensities $J_{\nu}$ are
substituted by the approximate radiation field Eq.\,\ref{ali} in order to
eliminate these variables from the solution vector Eq.\,\ref{psi1}. In
principle the ALO may be of arbitrary form as long as the iteration procedure
converges. In practice however an optimum choice is desired in order to achieve
convergence with a minimum amount of iteration steps.

In the case of diagonal (local) lambda operators the mean intensity $J_d$ at a
particular depth  $d$ in the current iteration step is computed solely from the
local source function $S_d$ and a correction term $\Delta J_d$, the latter
involving the source functions (of all depths) from the previous iteration.
Dropping the iteration count and introducing indices denoting depth points we
can rewrite Eq.\,\ref{ali}:
\begin{equation}
J_d=\Lambda^{\star}_{d,d}S_d+\Delta J_d.
\end{equation}
In the discrete form we now think of $\Lambda^{\star}$ as a matrix acting on a
vector whose elements comprise the source functions of all depths. Then
$\Lambda^{\star}_{d,d}$ is the diagonal element of the $\Lambda^{\star}$ matrix
corresponding to depth point $d$. Writing $\Lambda^{\star}_{d,d}\equiv B_d$
we have a purely local
expression for the mean intensity:
\begin{equation}
J_d=B_dS_d+\Delta J_d.
\end{equation}
Much better convergence is obtained if the mean intensity is computed not only
from the local source function but also from the source function of the
neighboring depths points. Then the matrix representation of $\Lambda^{\star}$
is of tri-diagonal form and we may write:
\begin{equation}\label{trij}
J_d=C_{d-1}S_{d-1}+B_dS_d+A_{d+1}S_{d+1}+\Delta J_d \ ,
\end{equation}
where $C_{d-1}$ and $A_{d+1}$ represent the upper and lower subdiagonal
elements of $\Lambda^{\star}$ and $S_{d\pm 1}$ the source functions at the
adjacent depths. We emphasize again that the actual source functions in
Eq.\,\ref{trij} are computed from the actual population densities and
temperature which are unknown. We therefore have a non-linear set of equations
which is solved by either a Newton-Raphson iteration or other techniques,
resulting in the solution of a tri-diagonal linear equation of the form
Eq.\,\ref{tri}. It was shown in Olson \etal (1986) that the elements of the
optimum $\Lambda^{\star}$ matrix are given by the corresponding elements of the
exact $\Lambda$ matrix.

\section{Solution of the non-linear equations by iteration}

The complete set of non-linear equations for a single iteration step
Eq.\,\ref{ali} comprises at each depth the equations for statistical,
radiative, and hydrostatic equilibrium and the particle conservation equation.
For the numerical solution we introduce discrete depth and frequency grids. The
equations are then linearized and solved by a suitable iterative scheme.
Explicit angle dependency of the radiation field is not required here and
consequently eliminated by the use of variable Eddington factors. Angle
dependency is only considered in the formal solution of the transfer equation.

\subsection{Discretization and Linearization}

A depth grid is set up and we start from a gray approximation computing a LTE
continuum model using the Uns\"old-Lucy temperature correction procedure. Depth
points (typical number is 90) are set equidistantly on a logarithmic
(Rosseland) optical depth scale. Our NLTE code uses the column mass as an
independent depth variable.

The frequency grid is established based upon the atomic data input file.
Frequency points are set blue- and redward of each absorption edge and for each
spectral line. Gaps are filled in by setting continuum points. Finally, the
quadrature weights are computed. Frequency integrals appearing e.g\@. in
Eq.\,\ref{combi} are replaced by quadrature sums and differential quotients
involving depth derivatives by difference quotients.

All variables $x$ are replaced by $x\rightarrow x+\delta x$ where $\delta x$
denotes a small perturbation of $x$. Terms not linear in these perturbations
are neglected. The perturbations are expressed by perturbations of the basic
variables:
\begin{equation}\label{deltax}
\delta x=      \frac{\partial x}{\partial T  }\delta T  +
               \frac{\partial x}{\partial n_e}\delta n_e+
               \frac{\partial x}{\partial N  }\delta N  +
               \frac{\partial x}{\partial n_H}\delta n_H+
\sum_{l=1}^{NL}\frac{\partial x}{\partial n_l}\delta n_l .
\end{equation}

\subsection{Newton-Raphson iteration and alternative methods}\label{kanto}

As described in Sect.\,\ref{eins2} the linearized equations have a tri-diagonal
block-matrix form, see Eq.\,\ref{tri}. Inversion of the grand matrix ($\equiv
{\bf T}$ sized $(NN\cdot ND) \times (NN\cdot ND)$, i.e.\ about $10^4\times 10^4$
in typical applications) is performed with a block-Gaussian elimination scheme,
which means that our iteration of the non-linear equations represents a
multi-dimensional Newton-Raphson method. The problem is structurally simplified
when explicit depth coupling is avoided by the use of a local ALO, however, the
numerical effort is not much reduced, because in both cases the main effort
lies with the inversion of matrices sized $NN\times NN$. The Newton-Raphson
iteration involves two numerically expensive steps, first setting up the
Jacobian (comprising {$\mbox{\boldmath$\alpha, \beta, \gamma$}$}) and then
inverting it. Additionally, the matrix inversions necessary to solve
Eq.\,\ref{tri}  limit their size to about $NN=250$ because otherwise numerical
accuracy is lost.

Two variants recently introduced in stellar atmosphere calculations are able to
improve both, numerical accuracy and, most of all, computational speed.
Broyden's (1965) variant belongs to the family of quasi-Newton methods and it
was first used in model atmosphere calculations in Dreizler \& Werner (1991),
Hamann \etal (1991) and Koesterke \etal (1992). It avoids the repeated set-up
of the Jacobian by the use of an update formula. On top of this, it also gives
an update formula for the {\em inverse} Jacobian. Another variant, the
Kantorovich method was introduced into model atmosphere calculations by Hubeny
\& Lanz (1992). This method simply keeps fixed the Jacobian during the
linearization cycle and it is surprisingly stable.

\section{NLTE metal line blanketing}

Despite the capacity increase for the  NLTE treatment of model atmosphere
problems by introducing the ALI method combined with pre-conditioning
techniques, the blanketing by millions of lines from the iron group elements
arising from transitions between some $10^5$ levels could only be attacked with
the help of statistical methods. These have been introduced into NLTE model
atmosphere work by Anderson (1989). At the outset, model atoms are constructed
by combining many thousands of levels into a relatively small number of
superlevels which can be treated with ALI (or other) methods. Then, in order to
reduce the computational effort, two approaches were developed which vastly
decrease the number of frequency points (and hence the number of transfer
equations to be solved) to describe properly the complex frequency dependence
of the opacity. These two approaches have their roots in LTE modeling
techniques, where for the same reason statistical methods are applied for the
opacity treatment: The Opacity Distribution Function (ODF) and Opacity Sampling
(OS) approaches. Both are based on the circumstance that the opacity (in the
LTE approximation) is a function of two only local thermodynamic quantities.
Roughly speaking, each opacity source can be written in terms of a population
density and a photon cross-section for the respective radiative transition:
{$\kappa_\nu \sim n_l \sigma_{lu}(\nu)$}. In LTE the population follows from
the Saha-Boltzmann equations, hence $n_l=n_l(n_e,T)$. The OS and ODF methods
use such pre-tabulated (on a very fine frequency mesh) $\kappa_\nu(n_e,T)$
during the model atmosphere calculations. The NLTE situation is more
complicated, because pre-tabulation of opacities is not useful. The population
densities at any depth now also depend explicitly on the radiation field (via
the rate equations which substitute the TE Saha-Boltzmann statistics) and thus
on the populations in each other depth of the atmosphere. As a consequence, the
OS and ODF methods are not applied to opacity tabulations, but on tabulations
of the photon cross-sections $\sigma(\nu)$. These do depend on local quantities
only, e.g\@. line broadening by Stark and Doppler effects is calculated from $T$
and $n_e$. In the NLTE case the cross-section takes over the role which the
opacity played in the LTE case. More details on this topic can be found in
Werner \& Dreizler (1999).

Our strategy is the following. Before any model atmosphere calculation is
started, the atomic data are prepared by constructing superlevels, and the
cross-sections for superlines. Then these cross-sections are either sampled  on
a coarse frequency grid or ODFs are constructed. These data are put into the
model atom which is read by the model code.

\section{Summary}

The construction of metal line blanketed models in hydrostatic and radiative
equilibrium under NLTE conditions was the last and long-standing problem of
classical model atmosphere theory and it is finally solved with a high degree
of sophistication. The essential milestones for this development, starting from
the pioneering work of Auer \& Mihalas (1969) are:

\begin{itemize}
\item Introduction of the Accelerated Lambda Iteration (ALI, or ``operator
splitting'' methods), based upon early work by Cannon (1973) and Scharmer
(1981). First ALI model atmospheres were constructed by Werner (1986).
\item Introduction of statistical approaches to treat the iron group elements
in NLTE by Anderson (1989).
\item Linear formulation of the statistical equations (Rybicki \& Hummer 1991, 
Hauschildt 1993).
\item Computation of atomic data by Kurucz (1991), by the Opacity Project
(Seaton \etal 1994) and the Iron Project (Hummer \etal 1993).
\end{itemize}

\begin{acknowledgements}
Stellar atmosphere work in T\"ubingen is supported by DFG and DLR. I 
thank for financial support from the conference organizers.
\end{acknowledgements}

\end{document}